# Observation of cation-specific critical behavior at the improper ferroelectric phase transition in $Gd_2(MoO_4)_3$

Inger-Emma Nylund, Maria Tsoutsouva, Tor Grande, Dennis Meier[a]

Department of Materials Science and Engineering, NTNU Norwegian University of Science and Technology, NO-7491 Trondheim, Norway

## Abstract

Gadolinium molybdate is a classical example of an improper ferroelectric and ferroelastic material. It is established that the spontaneous polarization arises as a secondary effect, induced by a structural instability in the paraelectric phase, which leads to a unit cell doubling and the formation of a polar axis. However, previous X-ray diffraction studies on gadolinium molybdate have been restricted by the limited ability to include the entire $2\theta$ range in the analysis, and thus, at atomic scale, much remains to be explored. By applying temperature dependent X-ray diffraction, we observe the transition from the paraelectric tetragonal phase to the orthorhombic ferroelectric phase. The ferroelastic strain is calculated based on the thermal evolution of the lattice parameters and Rietveld refinement of the temperature dependent data reveals that the displacement of different cations follows different critical behavior, providing new insight into the structural changes that drive the improper ferroelectricity in gadolinium molybdate.

## I. Introduction

Gadolinium molybdate ($Gd_2(MoO_4)_3$) is a classical example for improper ferroelectricity, which has been studied intensively for more than half a century [1]. Improper ferroelectrics are a special class of electrically order materials, where the primary symmetry breaking order parameter is not the electric polarization. Instead, the phase transition to the ferroelectric state is driven by a structural or magnetic instability, which breaks inversions symmetry and leads to the formation of a polar axis. Thus, different from *proper* ferroelectrics, such as $LiNbO_3$, $BaTiO_3$, $Pb(Zr_xTi_{1-x})O_3$, ferroelectricity in improper systems only occurs as a secondary effect, promoting unusual correlation phenomena between spin, charge and lattice degree of freedoms. In $TbMnO_3$ and $MnWO_4$, for example, the emergence of a magnetically induced polarization gives rise to pronounced magnetoelectric and magnetocapacitance effects[2]–[5], and in hexagonal manganites $RMnO_3$ ($R$ = Sc, Y, In, Dy to Lu) geometrically driven ferroelectricity causes the formation of unusual domain wall structures and topologically protected domains[6], [7].

In the model material $Gd_2(MoO_4)_3$, it is a structural instability in the paraelectric high-temperature phase that leads to a doubling of the unit cell volume and a distortion of the unit cell. The latter leads to the formation of a polar axis, which is then accompanied by a spontaneous polarization, making Gd$_2$(MoO$_4$)$_3$ a uniaxial improper ferroelectric. The induced polarization is oriented along the crystallographic $c$-axis indicated in Figure 1.

---

[a] Author to whom correspondence should be addressed: dennis.meier@ntnu.no

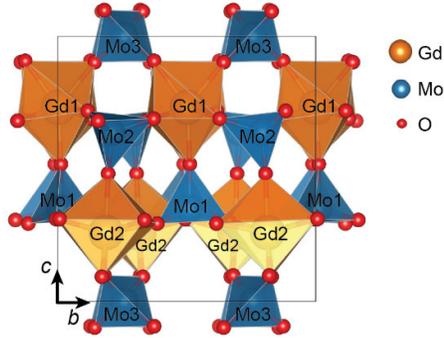

*Figure 1. Crystal structure of ferroelectric Gd$_2$(MoO$_4$)$_3$. The induced polarization is along the c-axis.*

It is established that the paraelectric-to-ferroelectric transition in Gd$_2$(MoO$_4$)$_3$ is a first order phase transition. This is supported by dielectric[8] and neutron scattering[9] measurements, in addition to the observation of latent heat[10] and lattice parameter hysteresis[11] at the phase transition. The structural primary order parameter that drives the phase transition has been determined as a phonon instability at the $(\frac{1}{2},\frac{1}{2},0)$ Brillouin zone corner of the tetragonal parent phase, leading to the unit cell doubling when this soft-mode "freezes in"[9]. Gd$_2$(MoO$_4$)$_3$ can thus be categorized as a geometric improper ferroelectric[12], similar to hexagonal manganites, which exhibit a unit cell tripling at $T_c$.

Ferroelastic switching can be achieved applying mechanical stress along the $b$-axis and results in an exchange of the $a$- and $b$-axis of the material [13]. Because of the improper nature of the ferroelectric order, however, switching the strain state inevitably results in polarization reversal, and vice versa by the application of an electric field, reflecting a one-to-one correlation between the ferroelastic and ferroelectric order [9]. Furthermore, Gd$_2$(MoO$_4$)$_3$ is birefringent[8], [14], and as a result, the material has found application as optical shutters, which are either mechanically or electrically controlled[15], and a potential use in memory devices has been proposed [15]. Thus, in addition to the intriguing physics of the coupled ferroelastic and ferroelectric order in Gd$_2$(MoO$_4$)$_3$, the material displays unique functional properties, being of interest for both fundamental and applied sciences.

Crystallographically, Gd$_2$(MoO$_4$)$_3$ can exist in two modifications referred to as the $\alpha$- and $\beta$-phase. The $\alpha$-phase is thermodynamically stable at room temperature and up to 857 °C, where it transforms to the stable high temperature $\beta$-phase. Gd$_2$(MoO$_4$)$_3$ can be retained in the metastable $\beta$-phase by cooling back below 857 °C, where transformation to the $\alpha$-phase is imperceptibly slow bellow 600 °C[16], [17].

$\beta$-Gd$_2$(MoO$_4$)$_3$ belongs to the tetragonal space group $P\bar{4}2_1m$ with $a$=7.39 and $c$=10.67 Å and two formula units per unit cell[18]. The unit cell can be said to have a layered structure which consists of (MoO$_4$)$^{2-}$-tetrahedra and Gd$^{3+}$ cations in seven-coordinated oxygen polyhedra[1]. In the $\beta$-phase, Gd$_2$(MoO$_4$)$_3$ is piezoelectric[9]. On cooling, $\beta$-Gd$_2$(MoO$_4$)$_3$ transforms at 159 °C into the improper ferroelastic[9], [19], improper ferroelectric[20] $\beta'$-phase. This is the phase transition at which the abovementioned unit cell doubling occurs, and the $\beta'$ phase belongs to the orthorhombic space group $Pba2$ with $a$=10.39, $b$=10.42 and $c$=10.70 Å[1], [18]. In addition to the unit cell doubling, the $P\bar{4}2_1m$ and $Pba2$ space groups are related through a 45° rotation about the collinear $c$-axis and a shift of the unit cell origin by (0, ½, 0). Thus, to describe the structural change between the $\beta$- and $\beta$'-phase, it is convenient to use the non-standard space group description $C\bar{4}2_1$ of the $\beta$ phase with $a$=10.46 and $c$=10.67 Å (and a shared unit cell origin) as it facilitates direct comparison of the two phases[18].

Here, we report on the structural evolution of $Gd_2(MoO_4)_3$ in the $Pba2$-phase by temperature dependent X-ray diffraction from ambient to above the ferroelectric phase transition. Going beyond previous studies of the thermal evolution of the lattice parameters [11], we investigate the temperature-driven displacement of the specific cations within the unit cell by applying Rietveld refinement of the diffraction data. The relative displacement of the cations in the improper ferroelectric and the paraelectric parent phase are obtained. The data reveals that the different cations follow different critical behavior, giving novel insight into the nature of the paraelectric-to-ferroelectric phase transition in $Gd_2(MoO_4)_3$.

## II. Method

### A. Synthesis

Single-phase powder of $Gd_2(MoO_4)_3$ was prepared via solid-state synthesis[21]. $MoO_3$ (99.97% trace metals basis, Sigma Aldrich) and $Gd_2O_3$ (99.9% trace metals basis, Sigma Aldrich) was dried for 5 h at 650 °C and 900 °C, respectively. Stoichiometric amounts were then mixed in an agate mortar with ethanol and dried. Pellets were pressed at 80 MPa and sintered at 950 °C for 25 h. The pellets were then crushed and finely ground in an agate mortar for XRD analysis.

### B. Characterization

In-situ temperature dependent X-ray diffraction experiments were performed, using a Bruker AXS, D8 Advanced diffractometer with Cu$K\alpha$ radiation ($\lambda$=0.154 nm), equipped with a VANTEC-1 position sensitive detector. The powder was dissolved in ethanol and deposited on a Pt-strip acting as the sample holder, as well as the heating source, using an MRI Physikalische Geräte GmbH high-temperature controller. The paraelectric-to-ferroelectric phase transition in $Gd_2(MoO_4)_3$ was investigated over an extended temperature range, from 25 °C to 145 °C every 5 °C, from 149 °C to 169 °C every 2 °C, and from 200 °C to 275 °C every 25 °C. Diffraction patterns were collected in the $2\theta$–range 7-110° with a step size of 0.016° and counting time of 1 s. A holding time of 5 minutes was used prior to each measurement which were performed at constant temperature. The temperature was calibrated against temperature dependent X-ray diffraction of a corundum standard powder deposited on the same Pt-strip, which gave an estimated temperature error less than 5 °C.

Rietveld refinement was performed with the fundamental parameters peak shape fitting, using TOPAS (v5). The improper ferroelectric and paraelectric phases were refined using the $Pba2$ and $P\bar{4}2_1m$ space groups, respectively, with starting parameters taken from Jeitschko[18]. The diffractograms from 159 °C and above were fitted using the $P\bar{4}2_1m$ space group. The refined parameters of the $P\bar{4}2_1m$ phase were transformed to $C\bar{4}2_1$ to facilitate direct comparison, as described above. The refinements were performed using a Chebychev polynomial background of order 7, and a preferred orientation factor for the (001) plane. In addition, the sample displacement, lattice parameters and cation positions were refined. The cations positions were refined without restrictions, with the exceptions indicated in Table I**Error! Reference source not found.**. The oxygen positions were not refined, and were fixed to the positions reported by Jeitschko[18].

Table I. Restrictions on cations during Rietveld refinement.

| Phase | Restriction |
|---|---|
| $Pba2$ | **Mo3**: z=0 |
| $P\bar{4}2_1m$ | **Mo1**, **Gd1**: y = -x+(3/2) |
|  | **Mo3**: x=1/2, y=1/2, z=0 |

## III. Results

### A. Lattice parameter evolution

The refined lattice parameters of $Gd_2(MoO_4)_3$ are presented as a function of temperature in Figure 2(a), and the volume evolution of the unit cell is shown in Figure 2(b). Visible in Figure 2(a), at the paraelectric-to-ferroelectric phase transition, is the expansion of the $c$-axis, and contraction and splitting of the tetragonal $a$-axis into the orthorhombic $a$- and $b$-axes below $T_c$=159 °C, which is in excellent agreement with literature[11]. A discontinuous volume change at the transition is evident in Figure 2(b), which reflects the first order nature of the phase transition, demonstrating that the synthesized powder is of high quality, as it reproduces results previously shown for single crystals [11].

The linear ($\alpha_a$, $\alpha_b$, $\alpha_c$) and volumetric ($\alpha_V$) thermal expansion coefficients (TEC) are determined for the $Pba2$ and $C\bar{4}2_1$ phases, respectively, and the values are presented in Table II. To avoid non-linear effects occurring near the phase transition, the TECs are calculated between 23 and 108 °C for the $Pba2$ phase, and between 201 and 276 °C for the $C\bar{4}2_1$ phase.

Table II. Linear ($\alpha_a$, $\alpha_b$, $\alpha_c$) and volumetric ($\alpha_V$) expansion coefficients of the $Pba2$ and $C\bar{4}2_1$ phase, between 23 and 108 °C and 201-276 °C, respectively.

| Phase | $\alpha_a$ [x10⁻⁵ °C⁻¹] | $\alpha_b$ [x10⁻⁵ °C⁻¹] | $\alpha_c$ [x10⁻⁵ °C⁻¹] | $\alpha_V$ [x10⁻⁵ °C⁻¹] |
|---|---|---|---|---|
| $Pba2$ | 2.68 | 1.61 | -0.924 | 3.36 |
| $C\bar{4}2_1$ | 0.611 | - | -0.029 | 1.15 |

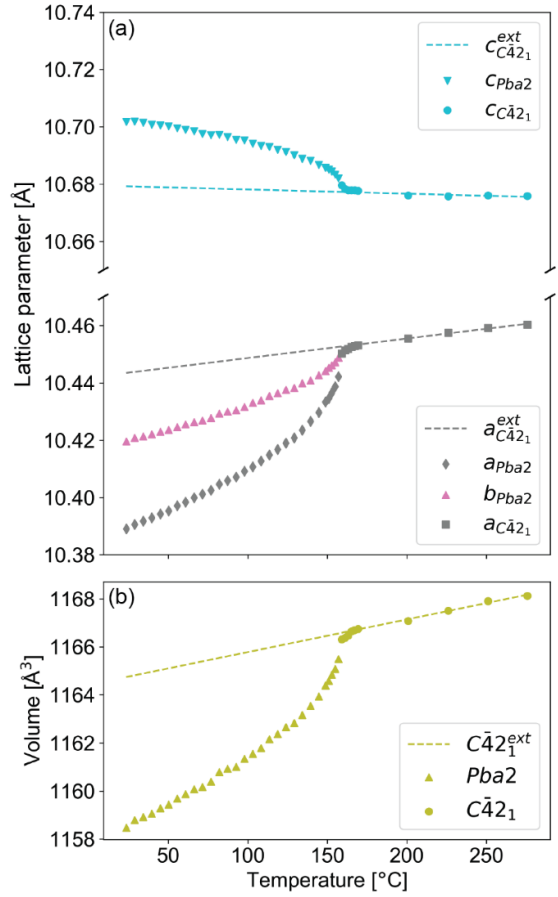

Figure 2. (a) Lattice parameter and (b) unit cell volume evolution as a function of temperature for $Gd_2(MoO_4)_3$ in the $Pba2$ and $C\bar{4}2_1$ phases. The calculated errors for the lattice parameters are 0.0004, 0.0004, and 0.0002 Å, or smaller for the a, b, and c parameters, respectively. The extrapolated a and c parameters, and the volume of the $C\bar{4}2_1$ phase are presented as stippled lines.

### B. Strain development

The temperature-dependent spontaneous strain $\epsilon_a(T)$ along the $a$-axis developing in the ferroic phase is defined as[22]

$$\epsilon_a(T) = \frac{a_{Pba2}(T) - a^{ext}_{C\bar{4}2_1}(T)}{a^{ext}_{C\bar{4}2_1}(T)}, \qquad (1)$$

where $a^{ext}_{C\bar{4}2_1}(T)$ is the extrapolated $a$ lattice parameter of the tetragonal $C\bar{4}2_1$ phase, and $a_{Pba2}(T)$ is the $a$ lattice parameter of the orthorhombic phase. The strain $\epsilon_b(T)$ along the $b$-axis and $\epsilon_c(T)$ along the $c$-axis are defined in an equivalent manner. The calculated strains $\epsilon_a(T)$, $\epsilon_b(T)$, and $\epsilon_c(T)$ are presented in Figure 3. The strains $\epsilon_a(T)$ and $\epsilon_b(T)$ along the $a$- and $b$-axes are negative, indicating that the $a$- and $b$-axes are compressed in the ordered state, whereas $\epsilon_c(T)$ is positive, showing an expansion of the $c$-axis in the $Pba2$ phase.

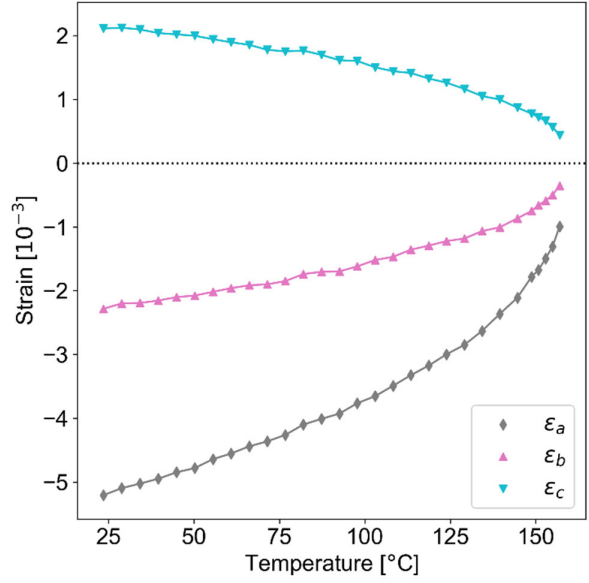

*Figure 3. Temperature evolution of the spontaneous strains along the a- ($\epsilon_a$), b- ($\epsilon_b$), and c-axes ($\epsilon_c$), respectively.*

### C. Relative displacement of atomic positions

The position of an atom in the basis of a unit cell is described by a vector $\vec{r} = (x\hat{a} + y\hat{b} + z\hat{c})$, where $0 < (x, y, z) \leq 1$ and $\hat{a}, \hat{b},$ and $\hat{c}$ are unit vectors along the three crystallographic axes. The positions of the cations in the $Pba2$ phase with the largest displacement relative to the $C\bar{4}2_1$ phase are plotted as a function of temperature in Figure 4(a) and Figure 4(b). (A full description of all atom positions in the $Pba2$ and $C\bar{4}2_1$ phase at 23 and 201 °C is presented in the Supplementary Material). The relative atomic displacements are defined as $\Delta\vec{r} = \vec{r} - \vec{r}'$, where $\vec{r}$ is the position of the atom in the unit cell of the $C\bar{4}2_1$ phase, and $\vec{r}'$ is the position of the atom in the $Pba2$ phase. The unit cell of Gd$_2$(MoO$_4$)$_3$ viewed along the $a$-axis is illustrated in Figure 4(c), showing the layered structure of the material. The cations which shift the most are located in layer 1 and layer 2, which are shown projected along the $c$-axis in Figure 4(d) and Figure 4(e), respectively.

Figure 4(a) and Figure 4(b) reveal that the cation shifts follow one of two qualitatively different behaviors: The cations either shift linearly as $T_c$ is approached, or the cation shifts display a discontinuity at the phase transition, following a non-linear trend. The linear fit is performed such that 0 displacement is reached at $T_c$, and the stippled lines indicating the non-linear trend is meant as a guide to the eyes. Interestingly, Mo3, which sits in layer 1 (Figure 4(a, d)), follows the linear trend in the $y$-direction and a non-linear trend in the $x$-direction. In layer 2, Gd2 shifts linearly in the $x$-direction, whereas Mo1 follows the non-linear behavior in the $y$-direction.

*Table III. Summary of the direction and the behavior of the cation displacements.*

| Layer | Cation | Direction | Behavior |
|---|---|---|---|
| 1 | Mo3 | $x$ | Non-linear |
| 1 | Mo3 | $y$ | Linear |
| 2 | Gd2 | $x$ | Linear |
| 2 | Mo1 | $y$ | Non-linear |

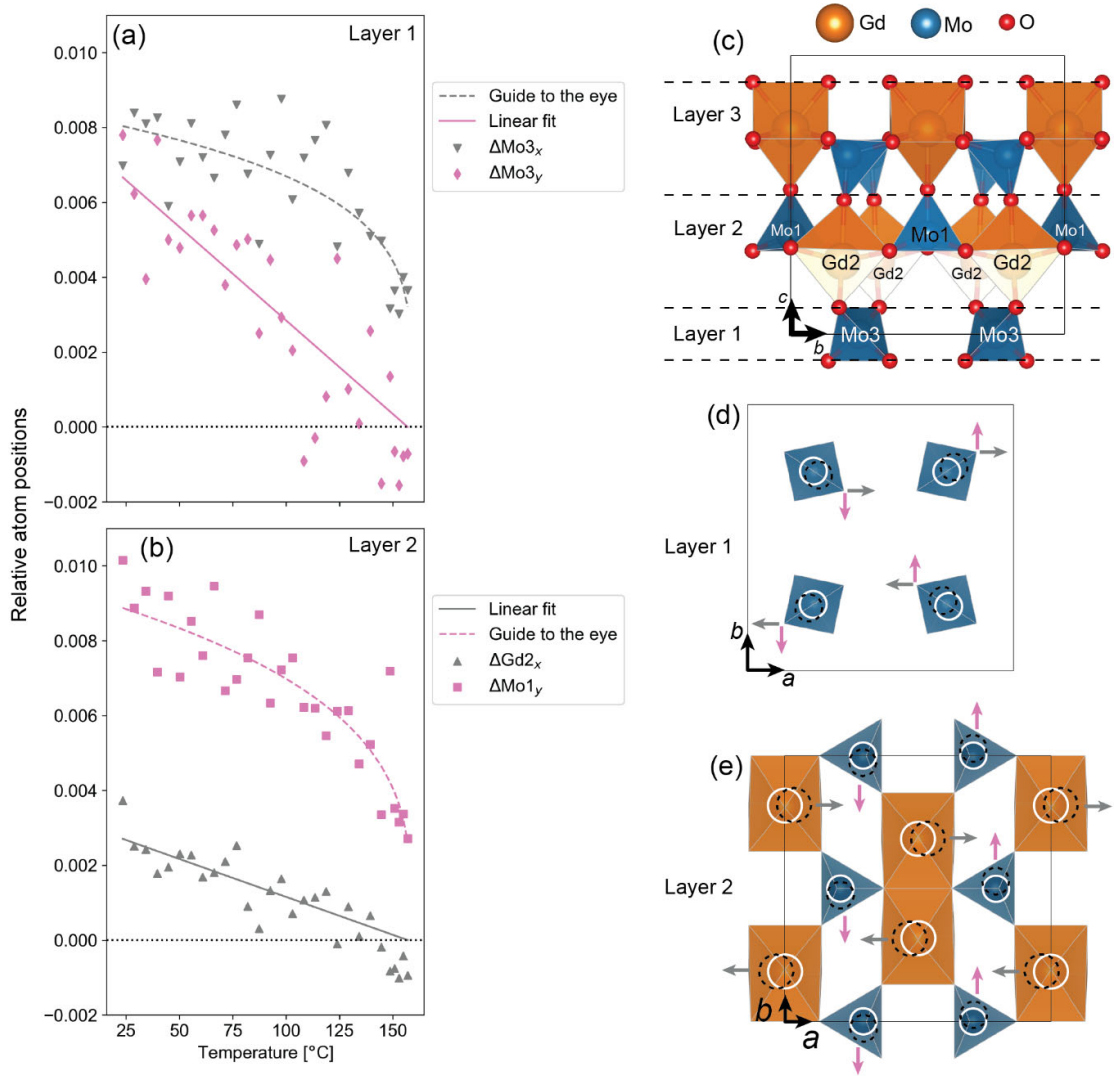

Figure 4. The temperature evolution of the relative displacement of Mo1, Mo3, and Gd2 in the $Pba2$ phase. (a) and (b) shows the relative atom positions as a function of temperature, determined from Rietveld refinement. (c), (d) and (e) show the $C\bar{4}2_1$ phase of GMO (with atom designation referring to the $Pba2$ phase). (c) GMO viewed as a three-layered structure when projected along the $a$-axis, and (d) and (e) show layer 1 and layer 2 viewed along the $c$-axis. The direction of the movement of the cation positions from the $C\bar{4}2_1$ phase to the $Pba2$ is indicated with white rings as the starting positions, and black dashed rings and arrows indicate the direction of the cation displacement. (The relative movement is exaggerated in the illustration, to make the displacement clearer). The relative displacement of oxygen anions accompanying the displacement of cations is not shown for simplicity.

## IV. Discussion

### A. Lattice parameters and strains

The discontinuity of the volume evolution at 159 °C (Figure 2) observed in our powder samples demonstrates the first order nature of the paraelectric-to-ferroelectric phase transition, consistent with the previous work on $Gd_2(MoO_4)_3$ single crystals by Kobayashi et al.[11]. We find that at the level of the individual cations, this first order phase transitions manifests as a step-like change in the displacement of Mo3 in the $x$-direction and Mo1 in the $y$-direction (Table III, Figure 4(a, b)).

Our calculations show that, qualitatively, the lattice strain evolution ($\epsilon_a, \epsilon_b$, and $\epsilon_c$) in Gd$_2$(MoO$_4$)$_3$ single-phase powder (Figure 3) and single crystals[11] exhibit similar behavior. Checking the values for all the strains at 50 °C and 150 °C gives a maximum deviation of about 25 % for $\epsilon_c$, and a maximum deviation of about 15 % for $\epsilon_a$ and $\epsilon_b$ between the data obtained on single-phase powder and single crystals. It is important to note, however, that our powder results are gained from Rietveld refinement, i.e., the whole diffractogram is taken into account in the determination of the lattice parameters and the calculated strain, whereas only specific diffraction peaks was considered to determine the lattice parameters from single crystals in the study by Kobayashi *et al.*[11].

### B. Cation displacements

Rietveld refinement is utilized to study the displacement of cations from their positions in the high temperature $C\bar{4}2_1$-phase relative to their positions in the ferroic $Pba2$-phase as a function of temperature (Figure 4). The oxygen atom positions are not refined in this work, and are fixed to the positions reported by Jeitschko[18].

#### Layer 1

In layer 1 in the $C\bar{4}2_1$-phase, the Mo3 tetrahedra are regular tetrahedra, meaning that all the distances between Mo3 and O-atoms are the same (Mo3-O9 = 1.7319Å). As the material is cooled down, the Mo3 shifts within the $xy$-plane of the $Pba2$-phase, shifting linearly with temperature in the $y$-direction and non-linearly along the $x$-direction. This displacement leads to two tetrahedra positioned closer to the center of the unit cell and two tetrahedra shifted towards the corners of the unit cell, as illustrated in Figure 4(d). The Mo3 remains at z=0 before and after the transition. At room temperature, the tetrahedra are no longer regular, but distorted, with a 1.1% difference between the longest and shortest Mo-O distances, i.e., the Mo3-tetrahedra in layer 1 are shifted and slightly distorted upon transitioning into the improper ferroelectric phase.

Comparing the different magnitude of the strains with the displacement of the Mo3, we see that the strain is largest along the $a$-axis of the unit cell, i.e. along the same axis as the non-linear displacement of the Mo3. At room temperature, however, the total magnitude of the relative shift of the Mo3 cation is about 0.008 in both directions. The special orientation of the Mo3-tetrahedra in layer 1 is such that an edge faces upwards along the $+z$-direction (and likewise an edge also faces downwards along $-z$).

#### Layer 2

Layer 2 consists of both Mo1-tetrahedra and Gd2-polyhedra, where each Gd-atom is bonded to seven oxygen atoms. For both Mo1 and Gd2, the shift along $z$ is minimal after the phase transition. As demonstrated in Figure 4(b), the major shifts of cations in this layer occur for Mo1 along the $y$-axis and Gd2 along the $x$-axis. Noteworthy for the polyhedra and tetrahedra in this layer is that they all have an apex pointing along the $+z$-direction. Furthermore, the tetrahedra are all corner sharing with the Gd2-polyhedra within layer 2 and share their apex-oxygen with the apex-oxygen, which points along the $-z$-direction of the Gd2-polyhedra of layer 3. The Gd2-polyhedra in layer 2 are corner sharing with Mo-tetrahedra in all the layers and, in addition, two and two Gd2-polyhedra are edge sharing within layer 2.

At the phase transition, the Mo1 shows a step-like displacement along the $y$-axis and follows a non-linear displacement towards room temperature. The shift of the Mo1-cation leads to a rotation of the Mo1-tetrahedra. The tetrahedra at $y\sim0$ rotate clockwise and the tetrahedra at $y\sim0.5$ rotate counterclockwise. In the $C\bar{4}2_1$-phase, the average distance Mo-O=1.7495 Å, with a 3.6 % deviation between the longest and shortest distance. In the $Pba2$-phase, the average distance is 1.7597 Å, with a 4.2 % difference between the longest and shortest bond length. The total shift of the relative position from the $C\bar{4}2_1$-phase to room temperature is about 0.009, which is similar to the Mo3 relative shift in layer 1. The displacement of Gd2 is smaller than the other relative cation displacements reported here. However, it has a clear linear trend with temperature away from the high temperature position. The average Gd-O distance in the $C\bar{4}2_1$-phase is 2.3535 Å, with a 5.6 % difference between the shortest and longest distance. At room temperature, the average distance is 2.3550 Å, with a 10.1 % difference. This clearly demonstrates that the Gd-polyhedra in general are more distorted. In particular, they are more distorted by the phase transition than the Mo-tetrahedra. This can be explained by the fact that the $MoO_4^{2-}$-tetrahedra is a much more rigid structure than the seven coordinated Gd-polyhedra. The rigidity is proposed to be due to the more covalent character of the Mo-O bond relative to the Gd-O bond.

The cation displacement in layer 2 can be said to be opposite of the displacement in layer 1. Meaning that the discontinuous (and largest) cation displacement in layer 2 happens along the $y$-direction, as compared to layer 1, where the discontinuous displacement happens along the $x$-direction. In addition, this is opposite to the strain behavior, which is largest along the $a$-axis.

### Layer 3

The displacement of the cations in Layer 3 due to the phase transitions is very small and, hence, their deviation from the high temperature phase is not discussed in further detail. However, as the cations in layer 2 shift, so does the oxygen atoms which are connected to the cations in Layer 1 and Layer 3. The latter leads to a rotation of the Mo-tetrahedra in layer 3, such that the tetrahedra at $x\sim0$ rotate counterclockwise and the Mo-tetrahedra at $x\sim0.5$ rotate clockwise. The Gd-polyhedra in layer 3 are also slightly more distorted in the $Pba2$ phase compared to the $C\bar{4}2_1$ phase.

## V. Conclusion

The phase transition of single-phase $Gd_2(MoO_4)_3$ synthesized via the solid-state method was studied by non-ambient X-ray diffraction. Lattice parameters, volume, and strain evolution was determined as a function of temperature. A discontinuous volume evolution at the phase transition was observed, reflecting a first order phase transition, consistent with previous investigations performed on $Gd_2(MoO_4)_3$ single crystals[11]. Furthermore, a similar strain evolution was demonstrated. Rietveld refinement was performed to shed light on the evolution of the atomic displacements that occur at the improper ferroelectric phase transition by studying the individual cation displacements from the high temperature positions. The Mo3 atoms in layer 1 were found to shift along the $x$- and $y$-direction. Further, they follow the same trend as the macroscopic strain, in the sense that the initial discontinuous and the largest shift occurred in the same direction as the largest strain. At room temperature, however, the total displacement was about equal in both the $x$- and $y$-direction. In layer 2, Mo1 was demonstrated to shift the most and non-linearly along the $y$-direction. Gd2, in layer 2, showed a clear linear displacement with temperature along the $x$-direction, even though the total displacement was smaller than for the other cations. Thus, the cation movement in layer 2 can be said to have an "opposite" trend to the cation movement in layer 1 and the spontaneous strain. This study demonstrates that individual cations in $Gd_2(MoO_4)_3$ follow two qualitatively different behaviors upon transitioning into the improper ferroelectric phase. The latter represents a so far hidden degree of freedom, providing additional insight into the microscopic origin of the spontaneous polarization in gadolinium molybdate and the complexity of structural phase transitions in improper ferroelectric materials in general.

## Supplementary Material

See Supplemental Material for details on the temperature calibration and the complete results of the Rietveld refinement of the structure at 23 °C and 201 °C.

## Availability of data

The data that support the findings of this study are available from the corresponding author upon reasonable request.

## Acknowledgements

Frida Paulsen Danmo is acknowledged for helping with the solid-state synthesis, and Ola G. Grendal for fruitful discussions regarding Rietveld refinement. The Research Council of Norway is acknowledged for financial support through the Projects FASTS (No. 250403/F20) and BORNIT (275139/F20).

## Authors' contributions

TG and DM initiated the project. IEN and MT performed XRD experiment and analysis with supervision from TG. IEN synthesized the powder. IEN wrote the manuscript under supervision of TG and DM, with contributions from all the authors.

# Supplementary Material: Observation of cation-specific critical behavior at the improper ferroelectric phase transition in $Gd_2(MoO_4)_3$

Inger-Emma Nylund, Maria Tsoutsouva, Tor Grande, Dennis Meier[a]

Department of Materials Science and Engineering, NTNU Norwegian University of Science and Technology, NO-7491 Trondheim, Norway

## Results from Rietveld refinement

Details on how the Rietveld refinement was performed is given in the main text. Table I and Table II show the refined cation positions (Gd and Mo), and oxygen positions reported by Jeitschko[1] in the $Pba2$ and $C\bar{4}2_1$ phase, at 23 and 201 °C, respectively. Below is a short note on the transformation from the non-standard space group $C\bar{4}2_1$ to the standard space group $P\bar{4}2_1m$, used in the Rietveld refinement.

$Pba2$
$a = 10.3891$ Å
$b = 10.4196$ Å
$c = 10.7018$ Å

*Table I. Atom positions in the $Pba2$ phase at 23 °C.*

| Atom  | x      | y      | z      |
|-------|--------|--------|--------|
| Gd(1) | 0.1873 | 0.4975 | 0.7330 |
| Gd(2) | 0.4963 | 0.3131 | 0.2600 |
| Mo(1) | 0.2063 | 0.4899 | 0.3517 |
| Mo(2) | 0.0029 | 0.2070 | 0.6380 |
| Mo(3) | 0.2430 | 0.2422 | 0.0    |
| O(1)  | 0.1921 | 0.4882 | 0.5186 |
| O(2)  | 0.4801 | 0.3053 | 0.4825 |
| O(3)  | 0.1287 | 0.0069 | 0.3112 |
| O(4)  | 0.4940 | 0.1280 | 0.6899 |
| O(5)  | 0.1579 | 0.1557 | 0.6815 |
| O(6)  | 0.1571 | 0.3360 | 0.3074 |
| O(7)  | 0.3840 | 0.3837 | 0.7191 |
| O(8)  | 0.3848 | 0.1145 | 0.2941 |
| O(9)  | 0.1255 | 0.1708 | 0.0937 |
| O(10) | 0.3174 | 0.1264 | 0.9074 |

[a] Author to whom correspondence should be addressed: dennis.meier@ntnu.no

| | | | |
|---|---|---|---|
| O(11) | 0.3545 | 0.3197 | 0.0984 |
| O(12) | 0.1704 | 0.3571 | 0.9024 |

$C\bar{4}2_1$
$a = 10.4555$ Å
$c = 10.6761$ Å

*Table II. Atom positions in the non-standard $C\bar{4}2_1$ phase at 201 °C.*

| Atom | x | y | z |
|---|---|---|---|
| Gd(1) | 0.1868 | 0.5 | 0.7368 |
| Mo(1) | 0.2065 | 0.5 | 0.3563 |
| Mo(3) | 0.25 | 0.25 | 0.0 |
| O(1) | 0.1952 | 0.5 | 0.5195 |
| O(3) | 0.1289 | 0.0 | 0.3109 |
| O(5) | 0.1389 | 0.1372 | 0.7005 |
| O(9) | 0.1377 | 0.1770 | 0.0955 |

## Transformation from $C\bar{4}2_1$ to $P\bar{4}2_1m$

The notation for describing the GMO crystal structure used here is adopted from Jeitschko[1], both in terms of cation nomenclature and utilizing the non-standard space group $C\bar{4}2_1$ to describe the high-temperature phase, to facilitate direct comparison between the two phases.

Given that all the atom positions in the $C\bar{4}2_1$ phase is described by $r = x + y + z$, where $0 \leq x, y, z < 1$. Then, in the $P\bar{4}2_1m$ phase, the atom positions can be described by $r' = x' + y' + z'$, $0 \leq x', y', z' < 1$, where:

$$x' = x + y,$$
$$y' = -x + y,$$
$$z' = z,$$

and the unit cell center must be additionally shifted by $[0, \frac{1}{2}, 0]$.